\documentstyle[amsfonts,aps]{revtex}

\begin{document}
\title{Alternative Formulation of The Quantum Electroweak Theory}
\author{Jun-Chen Su$^{1,2}$}
\address{1. Department of Physics, Harbin Institute of Technology, Harbin 150006,\\
People's Republic of China\\
2. Center for Theoretical Physics, College of Physics,\\
Jilin University, Changchun 130023, People's Republic of China}
\maketitle

\begin{abstract}
~~ The quantization of the electroweak theory is performed starting from the
Lagrangian given in the so-called unitary gauge in which the unphysical
Goldstone fields disappear. In such a Lagrangian, the unphysical
longitudinal components of the gauge fields and the residual gauge degrees
of freedom are naturally eliminated by introducing the Lorentz gauge
condition and the ghost equation. In this way, the quantum theory given in $%
\alpha $-gauge is perfectly established in the Lagangian formalism by the
Faddeev-Popov approach or the Lagrange multiplier method in the framework of
SU(2)$\times $U(1) gauge symmetry. The theory established is not only
simpler than the ordinary R$_\alpha -$gauge theory, but also explicitly
renormalizable. The unitarity of the S-matrix is ensured by the $\alpha -$%
limiting procedure proposed previously. Especially, it is shown that the
electroweak theory without involving the Higgs boson can equally be
formulated whitin the $SU(2)\times U(1)$ symmetry and exhibits good
renormalizability. The unitarity of such a theory may also be guaranteed by
the $\alpha $-limiting procedure.


\end{abstract}

\section{Introduction}

In our recent publication$^{[1]}$, it has been shown that the conventional
viewpoint that the massive gauge field theory can not be set up without
introducing the Higgs mechanism is not true. In fact, a certain massive
gauge field theories can be well established on the basis of
gauge-invariance principle without recourse to the Higgs mechanism. The
essential viewpoints to achieve this conclusion are: (a) a massive gauge
field\ must be viewed as a constrained system in the whole space of vector
potentials and the Lorentz condition, as a necessary constraint, must be
introduced from the beginning and imposed on the Lagrangian; (b) The
gauge-invariance\ of a gauge field theory should be generally examined from
the action of the field other than from the Lagrangian because the action is
of more fundamental dynamical meaning than the Lagrangian. Particularly, for
a constrained system such as the massive gauge field, the gauge-invariance
should be seen from its action given in the physical space defined by the
Lorentz condition. This concept is well-known in Mechanics; (c) In the
physical space, only infinitesimal gauge transformations are possibly
allowed and necessary to be considered. This fact was clarified originally
in Ref.[2]. Based on these points of view, it is easy to see that the
massive gauge field theory in which all the gauge bosons have the same
masses are surely gauge-invariant. Obviously, the Quantum chromodynamics
(QCD) with massive gluons fulfils this requirement because all the gluons
can be considered to have the same masses. As has been proved, such a QCD is
not only renormalizable, but also unitary$^{[3]}$. The renormalizability and
unitarity are warranted by the fact that the unphysical degrees of freedom
existing in the massive Yang-Mills Lagrangian are completely eliminated by
the introduced constraint conditions on the gauge field and the gauge group,
i.e. the Lorentz condition and the ghost equation. In the tree-diagram
level, as easily verified, the unphysical longitudinal part of the gauge
boson propagator gives a vanishing contribution to the S-matrix elements
because in the QCD Lagrangian, there only appear the vector currents of
quarks coupled to the gluon fields and in each current, the quarks and /or
the antiquarks are of the same flavor. For loop-diagrams, it has been proved
that the unphysical intermediate states are all cancelled out in S-matrix
elements. However, for the weak interaction, apart from the vector currents
of fermions, there appear the axial vector currents of fermions. In
particular, the charged and neutral gauge bosons as well as the charged and
neutral fermions are required to have different masses. In this case, it is
impossible to construct a gauge-invariant action without introducing the
Higgs mechanism. Within the framework of the Higgs mechanism, the
electro-weak-unified theory was successfully set up on the basis of the $%
SU(2)\times U(1)$ gauge-invariance$^{^{[4-6]}}$. The physical implication of
such a theory was originally revealed in the so-called unitary gauge in
which all the Goldstone fields are absent. The Lagrangian given in the
unitary gauge is obtained from the original $SU(2)\times U(1)$
gauge-symmetric Lagrangian which includes all the scalar fields in it by the
Higgs transformation.$^{[4-6]}$. Such a Lagrangian was initially used to
establish the quantum theory. The free massive gauge boson propagator
derived from the this theory is of the form$^{[4-8]}$%
\begin{equation}
iD_{\mu \nu }^j(k)=\frac{-i}{k^2-M_j^2+i\varepsilon }(g_{\mu \nu }-\frac{%
k_\mu k_\nu }{M_j^2})  \eqnum{1.1}
\end{equation}
where $j=W^{\pm }$ or $Z^0$. It is the prevailing point of view that the
above propagator explicitly ensures the unitarity of the S-matrix because
except for the physical pole at $k^2=M_j^2$, there are no other unphysical
poles to appear in the propagator. However, due to the bad ultraviolet
divergence of the second term in the propagator shown in Eq.(1.1), as
pointed out in the literature$^{[7]}$, Green's functions defined in the
unitary gauge theory are unrenormalizable. The uinrenormalizability of the
unitary gauge theory arises from the fact that the unphysical degrees of
freedom, i. e. the (four-dimensionally) longitudinal gauge fields and the
residual gauge degrees of freedom contained in the Lagrangian given in the
unitary gauge are not eliminated by introducing appropriate constraint
conditions. Later, the quantization of the electroweak theory was elegantly
carried out in the so-called $R_\alpha $- gauge by the Faddeev-Popov approach%
$^{[7,8]}$. In this quantization, the authors started from the original
Lagrangian which contains all the Goldstone fields in it and introduced the $%
R_\alpha $-gauge conditions$^{[7,8]}$ 
\begin{equation}
\partial ^\mu A_\mu ^a+\frac i2\alpha g(\phi ^{+}\tau ^aV-V^{+}\tau ^a\phi
)=0  \eqnum{1.2}
\end{equation}
\begin{equation}
\partial ^\mu B_\mu +\frac i2\alpha g^{\prime }(\phi ^{+}V-V^{+}\phi )=0 
\eqnum{1.3}
\end{equation}
where $A_\mu ^a$ and $B_\mu $ are the $SU(2)_T$ and $U(1)_Y$ gauge fields, $g
$ and $g^{\prime }$ are the $SU(2)_T$ and $U(1)_Y$ coupling constants,
respectively, $\alpha $ is the gauge parameter, $\tau ^a$ are the Pauli
matrices, 
\begin{equation}
\phi =\phi ^{\prime }+V  \eqnum{1.4}
\end{equation}
here 
\begin{equation}
\phi ^{\prime }=\frac 1{\sqrt{2}}\left( 
\begin{array}{c}
G_1+iG_2 \\ 
H+iG_0
\end{array}
\right)   \eqnum{1.5}
\end{equation}
in which $G_1,G_2$ and $G_0$ are the Goldstone fields and $H$ is the Higgs
field and 
\begin{equation}
V=\frac 1{\sqrt{2}}\left( 
\begin{array}{c}
0 \\ 
v
\end{array}
\right)   \eqnum{1.6}
\end{equation}
is the vacuum state doublet in which $v$ represents the vacuum expectation
value of the scalar field. The quantum theory built in the $R_{\alpha \text{ 
}}$ gauge has been widely accepted because the massive gauge boson
propagator given in this gauge is of the form$\ $ 
\begin{equation}
iD_{\mu \nu }^\alpha (k)=-i\left\{ \frac{g_{\mu \nu }-k_\mu k_\nu /k^2}{%
k^2-M_k^2+i\varepsilon }+\frac{\alpha k_\mu k_\nu /k^2}{k^2-\alpha
M_k^2+i\varepsilon }\right\}   \eqnum{1.7}
\end{equation}
With taking different values of the gauge parameter, we have different
propagators such as the ones given in the Landau gauge ($\alpha =0$), the 't
Hooft-Feynman gauge ($\alpha =1$) and the unitary gauge ($\alpha \rightarrow
\infty $), respectively. Since the above propagator shows good ultraviolet
behavior and therefore satisfies the power counting argument of
renormalizability, the quantum theory formulated in the $R_{\alpha \text{ }}$%
- gauge is considered to be renormalizable and there were some formal proofs
presented in the previous literature which seem to assert this point$%
^{[8-15]}$. Recently, however, H. Cheng and S. P. Li have presented a strong
argument which indicates that the quantum electroweak theory given in the $%
R_\alpha $-gauge is difficult to be renormalized, particularly, the
occurrence of double poles which are ultraviolet divergent renders the
multiplicative renormalization of the propagators to be impossible $^{[16]}$%
.Obviously, the difficulty of the renormalization of the $R_\alpha $-gauge
theory originates from the fact that the unphysical degrees of freedom
contained in the Lagrangian which was chosen to be the starting point of
quantization are not completely eliminated by the introduced $R_\alpha $-
gauge conditions. This point may clearly be seen from the Landau gauge in
which the $R_\alpha $- gauge conditions are reduced to the Lorentz gauge
conditions 
\begin{equation}
\partial ^\mu A_\mu ^a=0  \eqnum{1.8}
\end{equation}
\begin{equation}
\partial ^\mu B_\mu =0  \eqnum{1.9}
\end{equation}
These conditions imply vanishing of the longitudinal fields, $A_{L\mu }=0$
and $B_{L\mu }=0$ . But, the unphysical Goldstone fields could not be
constrained by the above constraint conditions. They are still remained in
the Lagrangian and play an important role in perturbative calculations (see
the illustration in Appendix A).

In this paper, we attempt to propose an alternative formulation of the
quantum electroweak theory in which the unphysical Goldstone bosons and even
the Higgs particle do not appear. According to the general principle of
constructing a renormalizable quantum field theory for a constrained system
such as the massive and massless gauge fields, the unphysical degrees of
freedom appearing in the Lagrangian ought to be all eliminated by
introducing necessary constraint conditions$^{[1,2]}$. This suggests that
the quantization of the electroweak theory may suitably be performed
starting from the Lagrangian given in the unitary gauge$^{[4,5]}$. This
Lagrangian originally was considered to be physical because the unphysical
Goldstone fields disappear in it. However, in such a Lagrangian still exist
the longitudinal components of the gauge fields and the residual gauge
degrees of freedom. These unphysical degrees of freedom may completely be
removed by introducing the Lorentz gauge conditions shown in Eqs.(1.8) and
(1.9) and the constraint condition on the gauge group (the ghost equation).
In this way, the quantum electroweak theory given in $\alpha $-gauge may
perfectly be set up by applying the Faddeev-Popov approach$^{[2]}$ or the
Lagrange undetermined multiplier method$^{[1]}$. In such a quantum theory,
the massive gauge boson propagators are still of the form as denoted in
Eq.(1.7) and hence exhibit explicit renormalizability of the theory.

There are two questions one may ask: One is that what is the gauge symmetry
of the theory established from the Lagrangian given in the unitary gauge ?
Another is how to ensure the unitarity of S-matrix elements evaluated from
such a theory? For the first question, we would like to mention that
ordinarily, the Lagrangian given in the unitary gauge, which is obtained
from the original $SU(2)\times U(1)$ symmetric Lagrangian by the Higgs
transformation, is explained to describe a spontaneously symmetry-broken
theory for which only the electric charge U(1)-symmetry is remained. This
concept comes from the assumption that the vacuum state in Eq.(1.6) is
invariant under gauge transformations. However, this concept is in
contradiction with the viewpoint adopted in the widely accepted quantum
theory established in Refs.[6-7] that the vacuum state in (1.6) is not
gauge-invariant, it undergoes the same gauge transformations as the scalar
field $\phi ^{\prime }$ shown in Eq.(1.5) does. Therefore, the present
quantum electroweak theory given in the $R_{\alpha \text{ }}$ gauge
quantization is set up from beginning to end on the basis of $SU(2)\times
U(1)$ gauge-symmetry and, as pointed out by t'Hooft$^{[17]}$, there actually
is no spontaneous symmetry-breaking to appear in such a theory. The
viewpoint mentioned above is essential. It enables us to build up a correct
electroweak theory from the Lagrangian given in the unitary gauge which is
still of the $SU(2)\times U(1)$ gauge-symmetry. Otherwise, one could not
build a reasonable quantum theory from the unitary gauge Lagrangian for the
electroweak-interacting system. To answer the second question, it is noted
that the propagator in Eq.(1.1) is derived from the on mass-shell vector
potential, while the propagator in Eq.(1.7) is derived from the off
mass-shell vector potential (see the explanation in the Appendix B). The
latter propagator may be derived from the generating functional of Green's
functions which is given In the path-integral quantization. This propagator
is suitable for the renormalization of off-shell Green's functions and also
for the renormalization of on-shell S-matrix elements because the
propagators appearing in the loop diagrams of S-matrix elements are
off-mass-shell even though the external momenta of the S-matrix elements are
on the mass-shell. Clearly, the propagators in Eq.(1.7) which is given in
the $\alpha $-gauge can be viewed as a parametrization of the propagators
given in the unitary gauge since in the limit: $\alpha \rightarrow \infty $,
the former propagator is converted to the latter one. Therefore, calculation
of a physical quantity may safely be done in the $\alpha $-gauge and then,
as was proposed and demonstrated previously$^{[18,19]}$, the $\alpha -$%
limiting procedure is necessary to be taken in the final step of the
calculation of a S-matrix element.

Furthermore, we try to build a theory in which the Higgs boson is removed.
This can be done by the requirement that the scalar field $\phi $ defined in
Eq.(1.4), which is a vector in the four-dimensional functional space, is
limited to the subspace in which the magnitude of the vector is equal to its
vacuum expectation value $v.$ Since the vacuum state in Eq.(1.5) obeys the
gauge-transformation law as the same as the scalar field $\phi $ does, it
will be shown that the action without involving the Higgs boson in it is
still of $SU(2)\times U(1)$ gauge-symmetry under the introduced Lorentz
condition. Therefore, the quantum electroweak theory without involving the
Higgs boson may also be set up within the framework of $SU(2)\times U(1)$
gauge-symmetry. The gauge boson propagators derived from such a theory is
still represented in Eq.(1.7). Therefore, the renormalizability of such a
theory is no problems and its unitarity can also be guaranteed by the $%
\alpha $-limiting procedure.

The rest of this paper is arranged as follows. In Sect.2, we describe the
quantization based on the Lagrangian given in the unitary gauge and the
Lorentz gauge condition. In Sect.3, some Ward-Takahashi identities$^{[20]}$
are derived. Sect.4 serves to inclusion of the quarks. Sect.4 is used to
formulate the theory without involving the Higgs boson. In the last section,
some remarks are presented. In Appendix A, we take an example to illustrate
the role of Goldstone particles in the ordinary $R_\alpha $-gauge theory.
Appendix B is used to explain the difference and the relation between the
both propagators derived in the unitary gauge and the $\alpha $-gauge.

\section{Quantization}

In this section, we describe the quantization of the electroweak theory
based on the Lagrangian given in the unitary gauge. For one generation of
leptons , the Lagrangian is$^{[4,5]}$%
\begin{equation}
{\cal L}={\cal L}_g+{\cal L}_f+{\cal L}_\phi  \eqnum{2.1}
\end{equation}
where ${\cal L}_g,{\cal L}_f$ and ${\cal L}_\phi $ are the parts of the
Lagrangian for the gauge fields, the lepton fields and the scalar fields,
respectively. They are written in the following 
\begin{equation}
{\cal L}_g=-\frac 14F^{a\mu \nu }F_{\mu \nu }^a-\frac 14B^{\mu \nu }B_{\mu
\nu }  \eqnum{2.2}
\end{equation}
where 
\begin{equation}
F_{\mu \nu }^a=\partial _\mu A_\nu ^a-\partial _\nu A_\mu ^a+g\epsilon
^{abc}A_\mu ^bA_\nu ^c  \eqnum{2.3}
\end{equation}
and 
\begin{equation}
B_{\mu \nu }=\partial _\mu B_\nu -\partial _\nu B_\mu  \eqnum{2.4}
\end{equation}

\begin{equation}
{\cal L}_f=\overline{L}i\gamma ^\mu D_\mu L+\overline{l}_Ri\gamma ^\mu D_\mu
l_R  \eqnum{2.5}
\end{equation}
where 
\begin{equation}
L=\left( 
\begin{array}{c}
\nu _L \\ 
l_L
\end{array}
\right)  \eqnum{2.6}
\end{equation}
is the doublet formed by a left-handed neutrino field $\nu _L$ and a
left-handed charged lepton field $l_l$, $l_R$ is the singlet of a
right-handed charged lepton field and 
\begin{equation}
D_\mu =\partial _\mu -ig\frac{\tau ^a}2A_\mu ^a-ig^{\prime }\frac Y2B_\mu 
\eqnum{2.7}
\end{equation}
is the covariant derivative in which $\frac{\tau ^a}2(a=1,2,3)$ and $\frac Y2
$ are the generators of SU(2)$_T$ and U(1)$_Y$ groups respectively. 
\begin{equation}
{\cal L}_\phi =(D^\mu \phi _0)^{+}(D_\mu \phi _0)-\mu ^2\phi _0^{+}\phi
_0-\lambda (\phi _0^{+}\phi _0)^2-f_l(\overline{L}\phi _0l_R+\overline{l}%
_R\phi _0^{+}L)  \eqnum{2.8}
\end{equation}
where 
\begin{equation}
\phi _0=\frac 1{\sqrt{2}}\left( 
\begin{array}{c}
0 \\ 
H+v
\end{array}
\right)  \eqnum{2.9}
\end{equation}
in which $v=\sqrt{-2\mu ^2/\lambda }.$ The $\phi _0$ is a special
configuration of the scalar fields which is connected with the field
configuration shown in Eq.(1.4) by a gauge transformation $\phi =U\phi _0$
where $U=\exp $\{$\frac i2(g\tau ^a\theta ^a+g^{\prime }\theta ^0)$\}.

As mentioned in the Introduction, in the Lagrangian written above still
exist the unphysical longitudinal parts of the gauge fields which are
necessary to be eliminated by introducing the Lorentz gauge conditions shown
in Eqs.(1.7) and (1.8). The necessity of introducing the Lorentz condition
in this case may also be seen from the $R_\alpha $-gauge condition. In fact,
considering that the conditions in Eqs.(1.2) and (1.3) should suit to any
field configuration, certainly, it is suitable for the field configuration
given in the unitary gauge. It is easy to verify that 
\begin{equation}
\phi _0^{+}\tau ^aV-V^{+}\tau ^a\phi _0=0  \eqnum{2.10}
\end{equation}
\begin{equation}
\phi _0^{+}V-V^{+}\phi _0=0  \eqnum{2.11}
\end{equation}
so that in the unitary gauge, the $R_\alpha $-gauge conditions is reduced to
the Lorentz gauge conditions which are now rewritten as 
\begin{equation}
\partial ^\mu A_\mu ^i=0  \eqnum{2.12}
\end{equation}
where we have set $A_\mu ^0\equiv B_\mu $ and let $i=0,1,2,3.$

Before performing the quantization of the electroweak theory starting from
the Lagrangian and the Lorentz condition described above by the
Faddeev-Popov method$^{[2]}$, it is at first stressed that the Lagrangian $%
{\cal L}(x)$ described above, as mentioned in Introduction, still has the $%
SU(2)\times U(1)$ gauge symmetry, unlike the ordinary concept that the
Lagrangian merely has the electric charge $U(1)$-symmetry. For the
Lagrangians ${\cal L}_g$and ${\cal L}_f$, it is clear that they are still $%
SU(2)\times U(1)$ gauge-symmetric in the unitary gauge. While, for the
Lagrangian ${\cal L}_\phi $, as can easily be verified, it also keeps
invariant under the following $SU(2)\times U(1$) gauge transformations: 
\begin{equation}
\delta \phi _0=\frac i2(g\tau ^a\theta ^a+g^{\prime }\theta ^0)\phi _0 
\eqnum{2.13}
\end{equation}
\begin{equation}
\delta L=\frac i2(g\tau ^a\theta ^a-g^{\prime }\theta ^0)L  \eqnum{2.14}
\end{equation}
\begin{equation}
\delta l_R=-ig^{\prime }\theta ^0l_R  \eqnum{2.15}
\end{equation}
\begin{equation}
\delta A_\mu ^i=D_\mu ^{ij}\theta ^j  \eqnum{2.16}
\end{equation}
where 
\begin{equation}
D_\mu ^{ij}=\delta ^{ij}\partial _\mu -g\varepsilon ^{ijk}A_\mu ^k 
\eqnum{2.17}
\end{equation}
In the above, the eigen-equations 
\begin{equation}
YL=-L,Yl_R=-2l_R,Y\phi _0=\phi _0,\tau ^al_R=0  \eqnum{2.18}
\end{equation}
and the definition 
\begin{equation}
\varepsilon ^{ijk}=\{ 
\begin{array}{c}
\epsilon ^{abc},\text{if }i,j,k=a,b,c=1,2,3; \\ 
0,\text{if }i,j\text{ and/or }k=0
\end{array}
\eqnum{2.19}
\end{equation}
have been used. Here, as said in Introduction, we adopt the concept implied
in the previous quantization of the electroweak theory carried out in the $%
R_\alpha $-gauge$^{^{[7,8]}}$ that the vacuum state shown in Eq.(1.6) is not
set to be gauge-invariant under the $SU(2)\times U(1)$ gauge
transformations. This vacuum state as well as the field $\phi _0$ shown in
Eq.(2.9) undergo the same gauge-transformation as the original scalar field $%
\phi $ denoted in Eq.(1.4) so that either the original Lagrangian ${\cal L}%
_\phi $ which includes the Goldstone fields in it or the Lagrangian ${\cal L}%
_\phi $ shown in Eq.(2.8) is still of the $SU(2)\times U(1)$ gauge symmetry.
Next, it is pointed out that to obtain a proper form of the ghost field
Lagrangian in the general $\alpha $-gauge, it is necessary to add the
identities in Eqs.(2.10) and (2.11) to the Lorentz condition in Eq.(2.12)
and write the constraint condition in a generalized form 
\begin{equation}
F^i[A,\phi _0]+\alpha \lambda ^i=0  \eqnum{2.20}
\end{equation}
where $\lambda ^i$ is an auxiliary function and 
\begin{equation}
F^i[A,\phi _0]=\partial ^\mu A_\mu ^i+\frac i2\alpha g^i(\phi _0^{+}\tau
^iV-V^{+}\tau ^i\phi _0)  \eqnum{2.21}
\end{equation}
here we have set $\tau ^0=1$ and $g^i=g,$ if $i=1,2,3$ and $g^i=g^{\prime },$
if $i=0$. This is because for the quantization performed in the Lagrangian
formalism, one has to make gauge transformations to the gauge condition
which will connect the Higgs field to other scalar fields.

Now we are in a position to carry out the quantization starting from the
Lagrangian given in the unitary gauge and the Lorentz condition. According
to the general procedure of the Faddeev-Popov approach of quantization$%
^{[2,21,22]}.$ we insert the following identity 
\begin{equation}
\Delta [A,\phi _0]\int D(g)\delta [F[A^g,\phi _0^g]+\alpha \lambda ]=1 
\eqnum{2.22}
\end{equation}
, where $g$ is an element of the $SU(2)\times U(1)$ group, into the
vacuum-to-vacuum transition amplitude, obtaining 
\begin{equation}
Z[0]=\frac 1N\int D\left( \Psi \right) D(g)\Delta [A,\phi _0]\delta
[F[A^g,\phi _0^g]+\alpha \lambda ]e^{i\int d^4x{\cal L}(x)}  \eqnum{2.23}
\end{equation}
where ${\cal L}(x)$ is the Lagrangian denoted in Eqs.(2.1)-(2.9) and $\Psi $
stands for all the field variables $(\overline{l},l,\overline{v},v,A_\mu
^a,B_\mu ,H)$ in the Lagrangian. Let us make a gauge transformation: $A_\mu
^i\rightarrow (A^{g^{-1}})_\mu ^i$ and $\phi _0\rightarrow \phi _0^{g^{-1}}$
to the functional in Eq.(2.23). Since the Lagrangian ${\cal L}(x)$ is
gauge-invariant and the functional $\Delta [A,\phi _0]$ as well as the
integration measure, as already proved in the literature$^{[2,8,15]}$, are
all gauge-invariant, the integral over the gauge group, as a constant, may
be factored out from the integral over the fields and put in the
normalization constant $N$. Thus, we have 
\begin{equation}
Z[0]=\frac 1N\int D(\Psi )\Delta [A,\phi _0]\delta [F[A,\phi _0]+\alpha
\lambda ]e^{i\int d^4\chi {\cal L}(x)}  \eqnum{2.24}
\end{equation}
The functional $\Delta [A,\phi _0]$ in the above, which may be evaluated
from the identity in Eq.(2.22) and the gauge-transformation shown in
Eqs.(2.13)-(2.17), will be expressed as$^{[2,21]}$ 
\begin{equation}
\Delta [A,\phi _0]=\det M[A,\phi _0]  \eqnum{2.25}
\end{equation}
where $M[A,\phi _0]$ is a matrix whose elements are 
\begin{equation}
M^{ij}(x,y)=\frac{\delta F_\theta ^i(x)}{\delta \theta ^j(y)}\mid _{\theta
=0}=\partial _x^\mu [D_\mu ^{ij}(x)\delta ^4(x-y)]+\frac i8\alpha
g^ig^j[V^{+}\tau ^i\tau ^j\phi _0(x)+\phi _0^{+}(x)\tau ^j\tau ^iV]\delta
^4(x-y)  \eqnum{2.26}
\end{equation}
Employing the familiar representation for the determinant$^{[2]}$ 
\begin{equation}
\det M=\int D(\overline{C},C)e^{i\int d^4xd^4y\overline{C}%
^i(x)M^{ij}(x,y)C^j(y)}  \eqnum{2.27}
\end{equation}
where $\overline{C}^i$ and $C^i$ are the mutually conjugate ghost field
variables, integrating Eq.(2.24) over the functions $\lambda ^i(x)$ with the
weight $exp[-\frac i2\alpha (\lambda ^i)^2]$ and then introducing the
external source terms for all the fields, we obtain from Eq.(2.24) the
generating functional of Green's functions such that 
\begin{equation}
Z[J]=\frac 1N\int D\left( \Phi \right) e^{i\int d^4x[{\cal L}_{eff}+J\cdot
\Phi )}  \eqnum{2.28}
\end{equation}
where $\Phi $ and $J$ designate respectively all the fields and external
sources including the ghosts and ${\cal L}_{eff}$ is the effective
Lagrangian for the system under consideration. With the following
definitions of the field variables 
\begin{equation}
W_\mu ^{\pm }=\frac 1{\sqrt{2}}(A_\mu ^1\mp iA_\mu ^2)  \eqnum{2.29}
\end{equation}
\begin{equation}
\left( 
\begin{array}{c}
Z_\mu \\ 
A_\mu
\end{array}
\right) =\left( 
\begin{array}{c}
\cos \theta _w-\sin \theta _w \\ 
\sin \theta _w\text{ }\cos \theta _w
\end{array}
\right) \left( 
\begin{array}{c}
A_\mu ^3 \\ 
B_\mu
\end{array}
\right)  \eqnum{2.30}
\end{equation}
\begin{equation}
C^{\pm }=\frac 1{\sqrt{2}}(C^1\mp iC^2),\overline{C}^{\pm }=\frac 1{\sqrt{2}}%
(\overline{C}^1\mp i\overline{C}^2)  \eqnum{2.31}
\end{equation}
and 
\begin{equation}
\left( 
\begin{array}{c}
C_z \\ 
C_\gamma
\end{array}
\right) =\left( 
\begin{array}{c}
\cos \theta _w-\sin \theta _w \\ 
\sin \theta _w\text{ }\cos \theta _w
\end{array}
\right) \left( 
\begin{array}{c}
C^3 \\ 
C^0
\end{array}
\right)  \eqnum{2.32}
\end{equation}
\begin{equation}
\left( 
\begin{array}{c}
\overline{C}_z \\ 
\overline{C}_\gamma
\end{array}
\right) =\left( 
\begin{array}{c}
\cos \theta _w-\sin \theta _w \\ 
\sin \theta _w\text{ }\cos \theta _w
\end{array}
\right) \left( 
\begin{array}{c}
\overline{C}^3 \\ 
\overline{C}^0
\end{array}
\right)  \eqnum{2.33}
\end{equation}
where $\theta _w$ is the Weinberg angle, the effective Lagrangian will be
represented as 
\begin{equation}
{\cal L}_{eff}={\cal L}_G+{\cal L}_F+{\cal L}_H+{\cal L}_{gf}+{\cal L}_{gh} 
\eqnum{2.34}
\end{equation}
where 
\begin{equation}
\begin{tabular}{l}
${\cal L}_G=-\frac 12W_{\mu \nu }^{+}W^{-\mu \nu }-\frac 14(Z^{\mu \nu
}Z_{\mu \nu }+A^{\mu \nu }A_{\mu \nu })+M_w^2W_\mu ^{+}W^{-\mu }+\frac 12%
M_z^2Z^\mu Z_\mu $ \\ 
$+ig[(W_{\mu \nu }^{+}W^{-\mu }-W_{\mu \nu }^{-}W^{+\mu })(\sin \theta
_wA^\nu +\cos \theta _wZ^\nu )+W_\mu ^{+}W_\nu ^{-}(\sin \theta _wA^{\mu \nu
}+\cos \theta _wZ^{\mu \nu })]$ \\ 
$+g^2\{W_\mu ^{+}W_\nu ^{-}(\sin \theta _wA^\mu +\cos \theta _wZ^\mu )(\sin
\theta _wA^\nu +\cos \theta _wZ^\nu )-W_\mu ^{+}W^{-\mu }(\sin \theta
_wA_\nu +\cos \theta _wZ_\nu )^2$ \\ 
$+\frac 12[(W_\mu ^{+})^2(W_\nu ^{-})^2-(W_\mu ^{+}W^{-\mu })^2]\}$%
\end{tabular}
\eqnum{2.35}
\end{equation}
in which 
\begin{equation}
W_{\mu \nu }^{\pm }=\partial _\mu W_\nu ^{\pm }-\partial _\nu W_\mu ^{\pm
},Z_{\mu \nu }=\partial _\mu Z_\nu -\partial _\nu Z_\mu ,A_{\mu \nu
}=\partial _\mu A_\nu -\partial _\nu A_\mu  \eqnum{2.36}
\end{equation}
and 
\begin{equation}
M_w=\frac 12gv,M_z=\frac{M_w}{\cos \theta _w}  \eqnum{2.37}
\end{equation}
\begin{equation}
{\cal L}_F=\overline{\nu }i\gamma ^\mu \frac 12(1-\gamma _5)\partial _\mu
\nu +\overline{l}(i\gamma ^\mu \partial _\mu -m_l)l+\frac g{\sqrt{2}}(j_\mu
^{-}W^{+\mu }+j_\mu ^{+}W^{-\mu })-ej_\mu ^{em}A^\mu +\frac e{\sin 2\theta _w%
}j_\mu ^0Z^\mu  \eqnum{2.38}
\end{equation}
in which 
\begin{equation}
j_\mu ^{-}=\overline{\nu }\gamma ^\mu \frac 12(1-\gamma _5)l=(j_\mu ^{+})^{+}
\eqnum{2.39}
\end{equation}
\begin{equation}
j_\mu ^{em}=\overline{l}\gamma _\mu l  \eqnum{2.40}
\end{equation}

\begin{equation}
j_\mu ^0=\overline{\nu }\gamma _\mu \frac 12(1-\gamma _5)\nu -\overline{l}%
\gamma _\mu \frac 12(1-\gamma _5)l+2\sin ^2\theta _wj_\mu ^{em}  \eqnum{2.41}
\end{equation}
and 
\begin{equation}
m_l=\frac 1{\sqrt{2}}f_lv  \eqnum{2.42}
\end{equation}
\begin{equation}
{\cal L}_H=\frac 12(\partial ^\mu H)^2-\frac 12m_H^2H^2+\frac g4(W^{+\mu
}W_\mu ^{-}+\frac 1{2\cos \theta _w}Z^\mu Z_\mu )(H^2+2vH)-\lambda vH^3-%
\frac \lambda 4H^4-\frac{f_l}{\sqrt{2}}\overline{l}lH  \eqnum{2.43}
\end{equation}
\begin{equation}
{\cal L}_{gf}=-\frac 1\alpha [\partial ^\mu W_\mu ^{+}\partial ^\nu W_\nu
^{-}+\frac 12(\partial ^\mu Z_\mu )^2+\frac 12(\partial ^\mu A_\mu )^2] 
\eqnum{2.44}
\end{equation}
and 
\begin{equation}
\begin{tabular}{l}
${\cal L}_{gh}=\overline{C}^{-}(\Box +\alpha M_w^2)C^{+}+\overline{C}%
^{+}(\Box +\alpha M_w^2)C^{-}+\overline{C}_z(\Box +\alpha M_z^2)C_z+%
\overline{C}_\gamma \Box C\gamma -ig\{(\partial ^\mu \overline{C}^{+}C^{-}$
\\ 
$-\partial ^\mu \overline{C}^{-}C^{+})(\cos \theta _wZ_\mu +\sin \theta
_wA_\mu )+(\partial ^\mu \overline{C}^{-}W_\mu ^{+}-\partial ^\mu \overline{C%
}^{+}W_\mu ^{-})(\cos \theta _wC_z+\sin \theta _wC_\gamma )$ \\ 
$+(\cos \theta _w\partial ^\mu \overline{C}_z+\sin \theta _w\partial ^\mu 
\overline{C}_\gamma )(C^{+}W_\mu ^{-}-C^{-}W_\mu ^{+})\}+\frac 12\alpha
gM_wH(\overline{C}^{+}C^{-}+\overline{C}^{-}C^{+}+\frac 1{\cos ^2\theta _w}%
\overline{C}_zC_z)$%
\end{tabular}
\eqnum{2.45}
\end{equation}
The external source terms in Eq.(2.28) are defined by 
\begin{equation}
\begin{tabular}{l}
$J\cdot \Psi =J^{-}\mu W^{+\mu }+J_\mu ^{+}W^{-\mu }+J_\mu ^zZ^\mu +J_\mu
^\gamma A^\mu +JH+\overline{\xi }_ll+\overline{l}\xi _l+\overline{\xi }_\nu
\nu +\overline{\nu }\xi _\nu $ \\ 
$+\overline{\eta }^{+}C^{-}+\overline{\eta }^{-}C^{+}+\overline{C}^{+}\eta
^{-}+\overline{C}^{-}\eta ^{+}+\overline{\eta }_zC_z+\overline{C}_z\eta _z+%
\overline{\eta }_\gamma C_\gamma +\overline{C}_\gamma \eta _\gamma $%
\end{tabular}
\eqnum{2.46}
\end{equation}
For the case of three generations of leptons, in Eq.(2.38) there will be the
kinetic energy terms for three neutrinos and three leptons, and the sums
over the number of generations of leptons should be included in
Eqs.(2.39)-(2.41). Correspondingly, the external source terms will be
extended to the case for three generations of leptons.

It is interesting to note that the quantized result shown above can be
obtained more directly by the Lagrange multiplier method$^{[1]}$. By this
method, we may incorporate the constraint condition in Eq.(2.20) into the
Lagrangian in Eq.(2.1) which is now extended to the form 
\begin{equation}
{\cal L}^{\prime }={\cal L-}\frac 12\alpha (\lambda ^i)^2  \eqnum{2.47}
\end{equation}
where ${\cal L}$ was written in Eqs.(2.1)-(2.9) and thus obtain a
generalized Lagrangian 
\begin{equation}
\begin{tabular}{l}
${\cal L}_\lambda ={\cal L}^{\prime }+{\cal \lambda }^iF^i[A,\phi _0]+\alpha
(\lambda ^i)^2$ \\ 
$={\cal L+\lambda }^iF^i[A,\phi _0]+\frac 12\alpha (\lambda ^i)^2$%
\end{tabular}
\eqnum{2.48}
\end{equation}
To construct a gauge-invariant theory, it is necessary to require the action
given by the above Lagrangian to be gauge-invariant under the gauge
transformations in Eqs.(2.13)-(2.16) and the constraint condition in
Eq.(2.20) 
\begin{equation}
\delta S_\lambda =\int d^4x\delta {\cal L}_\lambda =\int d^4x{\cal \lambda }%
^i(x)\{\partial _x^\mu [D_\mu ^{ij}(x)\theta ^j(x)]+\frac i8\alpha
g^ig^j[V^{+}\tau ^i\tau ^j\phi _0(x)+\phi _0^{+}(x)\tau ^j\tau ^iV]\theta
^j(x)\}=0  \eqnum{2.49}
\end{equation}
Since ${\cal \lambda }^i(x)\neq 0$, we have 
\begin{equation}
\partial _x^\mu [D_\mu ^{ij}(x)\theta ^j]+\frac i8\alpha g^ig^j[V^{+}\tau
^i\tau ^j\phi _0(x)+\phi _0^{+}(x)\tau ^j\tau ^iV]\theta ^j=0  \eqnum{2.50}
\end{equation}
This just is the constraint equations on the gauge group. Setting $\theta
^\alpha =\xi C^\alpha $ where $\xi $ is an anticommuting number and $%
C^\alpha $ are the ghost fields, we obtain from Eq.(2.50) the ghost
equations 
\begin{equation}
\partial _x^\mu [D_\mu ^{ij}(x)C^j]+\frac i8\alpha g^ig^j[V^{+}\tau ^i\tau
^j\phi _0(x)+\phi _0^{+}(x)\tau ^j\tau ^iV]C^j=0  \eqnum{2.51}
\end{equation}
Since these equations are the alternatives of the constraints on the gauge
group, they may also be incorporated into the Lagrangian ${\cal L}_\lambda $
by the Lagrange multiplier method. Thus, we have 
\begin{equation}
\begin{tabular}{l}
${\cal L}_\lambda ^{*}={\cal L+\lambda }^iF^i[A.\phi _0]+\frac 12\alpha
(\lambda ^i)^2+\overline{C}^i\{\partial _x^\mu [D_\mu ^{ij}(x)C^j]$ \\ 
$+\frac i8\alpha g^ig^j[V^{+}\tau ^i\tau ^j\phi _0(x)+\phi _0^{+}(x)\tau
^j\tau ^iV]C^j\}$%
\end{tabular}
\eqnum{2.52}
\end{equation}
where $\overline{C}^i$, acting as the Lagrange multipliers, are another kind
of ghost field variables which are conjugate to the variables $C^j$.

As we learn from the Lagrange multiplier method, all the variables in the
Lagrangian ${\cal L}_\lambda ^{*}$, including the dynamical variables, the
constrained variables and the Lagrange multipliers act as independent,
varying arbitrarily. Therefore, we may directly utilize the Lagrangian $%
{\cal L}_\lambda ^{*}$ to construct the generating functional of Green's
functions so as to achieve the final goal of quantization 
\begin{equation}
Z[J]=\frac 1N\int D\left( \Psi \right) D\left( \lambda \right) \exp \{i\int
d^4x({\cal L}_\lambda ^{*}+J\cdot \Psi )\}  \eqnum{2.53}
\end{equation}
where $\Psi $ stands for all the field variables but the Lagrange
multipliers and $J$ designates all the corresponding external sources. On
calculating the integral over $\lambda ^i,$ we precisely obtain the result
as shown in Eqs.( 2.28)-(2.46) which was given by the Faddeev-Popov approach.

In the end, we note that the effective action and the generating functional
obtained in this section, as easily proven, are invariant under a kind of
BRST- transformations$^{[21]}$. The BRST-transformations include the gauge
transformations shown in Eqs.(2.13)-(2.16) and the following transformations
for ghost fields 
\begin{equation}
\delta \overline{C}^i=\frac \lambda \alpha \partial ^\mu A_\mu ^i 
\eqnum{2.54}
\end{equation}
\begin{equation}
\delta C^i=-\frac \lambda 2g\varepsilon ^{ijk}C^jC^k  \eqnum{2.55}
\end{equation}
where $\lambda $ is an infinitesimal anticommuting number. Correspondingly,
the group parameters in Eqs.(2.13)-(2.16) should be represented by $\theta
^i=\lambda C^i.$ The BRST-invariance will leads to a set of Ward-Takahashi
identites$^{[20]}$ satisfied by the generating functionals as exhibited in
the next section.

\section{Ward-Takahashi identity}

In the preceding section, it was mentioned that the generating functional $%
Z[J]$ is invariant with respect to the BRST-transformations. the
BRST-transformations may be written as 
\begin{equation}
\delta \Phi _i=\lambda \triangle \Phi _i  \eqnum{3.1}
\end{equation}
where the $\triangle \Phi _i$ for every field can be explicitly written out
from Eqs.(2.13)-(2.16), (2.54) and (2.55). They are shown in the following 
\begin{equation}
\begin{tabular}{l}
$\triangle l=i\{\frac g{2\sqrt{2}}C^{-}(1-\gamma _5)\nu -e[C_\gamma +\frac 1{%
2\sin 2\theta _w}(1-\gamma _5-4\sin ^2\theta _w)C_z]l\}$ \\ 
$\triangle \overline{l}=-i\{\frac g{2\sqrt{2}}C^{+}\overline{\nu }(1+\gamma
_5)+e\overline{l}[C_\gamma +\frac 1{2\sin 2\theta _w}(1+\gamma _5-4\sin
^2\theta _w)C_z]\}$ \\ 
$\triangle \nu _L=i\frac g2[\frac 1{\cos \theta _w}C_z\nu _L+\sqrt{2}%
C^{+}l_L]$ \\ 
$\triangle \overline{\nu }_L=-i[\frac 1{\cos \theta _w}C_z\overline{\nu }_L+%
\sqrt{2}C^{-}\overline{l}_L]$ \\ 
$\triangle W_\mu ^{\pm }=\partial _\mu C^{\pm }\mp igC^{\pm }(\cos \theta
_wZ_\mu +\sin \theta _wA_\mu )\pm igW_\mu ^{\pm }(\cos \theta _wC_z+\sin
\theta _wC_\gamma )$ \\ 
$\triangle Z_\mu =\partial _\mu C_z-ig\cos \theta _w(C^{-}W_\mu
^{+}-C^{+}W_\mu ^{-})$ \\ 
$\triangle A_\mu =\partial _\mu C_\gamma -ig\sin \theta _w(C^{-}W_\mu
^{+}-C^{+}W_\mu ^{-})$ \\ 
$\triangle \overline{C}^{\pm }=\frac 1\alpha \partial ^\mu W_\mu ^{\pm }$ \\ 
$\triangle \overline{C}_z=\frac 1\alpha \partial ^\mu Z_\mu $ \\ 
$\triangle \overline{C}_\gamma =\frac 1\alpha \partial ^\mu A_\mu $ \\ 
$\triangle C^{\pm }=\pm i\sqrt{2}g(\cos \theta _wC_z+\sin \theta _wC_\gamma
)C^{\pm }$ \\ 
$\triangle C_z=ig\cos \theta _wC^{+}C^{-}$ \\ 
$\triangle C_\gamma =ig\sin \theta _wC^{+}C^{-}$ \\ 
$\triangle \varphi ^{\pm }=\pm \frac i2gC^{\pm }(H+v)$ \\ 
$\triangle \varphi ^0=-\frac g{2\cos \theta _w}C_z(H+v)$ \\ 
$\triangle H=0$%
\end{tabular}
\eqnum{3.2}
\end{equation}
The last three expressions which come from Eq.(2.13) indicate that in the
unitary gauge formulation of the theory, the $SU(2)\times U(1)$ gauge
transformation keeps the Higgs field to be invariant, while, creates three
Goldstone-type composite fields which consist of the Higgs field and the
ghost fields only without concerning Goldstone fields denoted in Eq.(1.5).
It is easy to prove that except for $\triangle \overline{C^{\pm }},\triangle 
\overline{C_z}$ and $\triangle \overline{C_\gamma }$, the other functions $%
\triangle \widetilde{\Phi }_i$ in Eq.(3.2) are nilpotent, $\delta \triangle 
\widetilde{\Phi }_i=0$ which means BRST-invariance of the functions $%
\triangle \widetilde{\Phi }_i.$

Let us define a generalized generating functional by including external
sources for the nilpotent functions $\triangle \widetilde{\Phi }_i$%
\begin{equation}
Z[J,K]=\frac 1N\int D(\Phi )\exp \{iS_{eff}+i\int d^4x[J_i(x)\Phi
_i(x)+K_i(x)\triangle \widetilde{\Phi }_i(x)]\}  \eqnum{3.3}
\end{equation}
where $J_i\Phi _i$ was shown in Eq.(2.46) and 
\begin{equation}
\begin{tabular}{l}
$K_i\triangle \widetilde{\Phi }_i=u_\mu ^{+}\triangle W^{-\mu }+u_\mu
^{-}\triangle W^{+\mu }+u_z^\mu \triangle Z_\mu +u_\gamma ^\mu \triangle
A_\mu +v^{+}\triangle C^{-}+v^{-}\triangle C^{+}+v_z\triangle C_z$ \\ 
$+v_\gamma \triangle C_\gamma +\overline{\chi }_l\triangle l+\triangle 
\overline{l}\chi _l+\overline{\chi }_\nu \triangle \nu _L+\triangle 
\overline{\nu }_L\chi _\nu +K^{+}\triangle \varphi ^{-}+K^{-}\triangle
\varphi ^{+}+\triangle \varphi ^0K^0$%
\end{tabular}
\eqnum{3.4}
\end{equation}
On making the BRST-transformation to the functional $Z[J,K]$ and noticing
the BRST-invariance of the functional, we obtain a W-T identity such that$%
^{[12,21,22]}$ 
\begin{equation}
\frac 1N\int D(\Phi )\int d^4x(\pm )J_i(x)\triangle \Phi _i(x)\exp
\{iS_{eff}+iJ\cdot \Phi +K\cdot \triangle \widetilde{\Phi }\}=0  \eqnum{3.5}
\end{equation}
where the signs ''$+"$ and ''$-$'' attribute to commuting and anticommuting
sources $J_i$ respectively. The above identity may be represented in terms
of differentials of $Z[J,K]$ with respect to the external sources. Here we
only write down specifically the identity satisfied by the generating
functional of connected Green's functions $W[J,K]$ which is defined by $%
Z=exp(iW)^{[19,21,22]}$ 
\begin{equation}
\begin{tabular}{l}
$\int d^4x\{\xi _l(x)\frac \delta {\delta \chi _l(x)}-\overline{\xi }_l(x)%
\frac \delta {\delta \overline{\chi }_l(x)}+\xi _\nu (x)\frac \delta {\delta
\chi _\nu (x)}-\overline{\xi }_\nu (x)\frac \delta {\delta \overline{\chi }%
_\nu (x)}+J_\mu ^{+}(x)\frac \delta {\delta u_\mu ^{+}(x)}+J_\mu ^{-}(x)%
\frac \delta {\delta u_\mu ^{-}(x)}$ \\ 
$+J_z^\mu (x)\frac \delta {\delta u_z^\mu (x)}+J_\gamma ^\mu (x)\frac \delta
{\delta u_\gamma ^\mu (x)}-\overline{\eta }^{+}(x)\frac \delta {\delta
v^{+}(x)}-\overline{\eta }^{-}(x)\frac \delta {\delta v^{-}(x)}-\overline{%
\eta }_z(x)\frac \delta {\delta v_z(x)}-\overline{\eta }_\gamma (x)\frac 
\delta {\delta v_\gamma (x)}$ \\ 
$+\frac 1\alpha \partial _\mu ^x\frac \delta {\delta J_\mu ^{+}(x)}\eta
^{+}(x)+\frac 1\alpha \partial _\mu ^x\frac \delta {\delta J_\mu ^{-}(x)}%
\eta ^{-}(x)+\frac 1\alpha \partial _x^\mu \frac \delta {\delta J_z^\mu (x)}%
\eta _z(x)+\frac 1\alpha \partial _x^\mu \frac \delta {\delta J_\gamma ^\mu
(x)}\eta _\gamma (x)\}W[J,K]$ \\ 
$=0$%
\end{tabular}
\eqnum{3.6}
\end{equation}

When we make a translation transformation: $\overline{C}^i\rightarrow 
\overline{C}^i+\overline{\lambda }^i$ to the functional $Z[J,K]$, then
differentiate the functional with respect to the $\overline{\lambda }^i(x)$
and finally set $\overline{\lambda }^i=0$, we get such a ghost equation that$%
^{[20-22]}$ 
\begin{equation}
\frac 1N\int D(\Phi )[\eta ^i(x)+\triangle F^i(x)]\exp \{\{iS_{eff}+iJ\cdot
\Phi +K\cdot \triangle \widetilde{\Phi }\}=0  \eqnum{3.7}
\end{equation}
where 
\begin{equation}
\triangle F^i(x)=\partial _x^\mu [D_\mu ^{ij}(x)C^j(x)]+\frac i8\alpha
g^ig^j[V^{+}\tau ^i\tau ^j\phi _0(x)+\phi _0^{+}(x)\tau ^j\tau ^iV]C^j(x) 
\eqnum{3.8}
\end{equation}
From the above equation, we may write out the following ghost equations via
the functional $W[J,K]$ 
\begin{equation}
\eta ^{+}(x)+\partial _\mu ^x\frac{\delta W}{\delta u_\mu ^{-}(x)}-i\alpha
M_w\frac{\delta W}{\delta K^{-}(x)}=0  \eqnum{3.9}
\end{equation}

\begin{equation}
\eta ^{-}(x)+\partial _\mu ^x\frac{\delta W}{\delta u_\mu ^{+}(x)}+i\alpha
M_w\frac{\delta W}{\delta K^{+}(x)}=0  \eqnum{3.10}
\end{equation}
\begin{equation}
\eta _z(x)+\partial _x^\mu \frac{\delta W}{\delta u_z^\mu (x)}+\alpha M_z%
\frac{\delta W}{\delta K^0(x)}=0  \eqnum{3.11}
\end{equation}

\begin{equation}
\eta _\gamma (x)+\partial _x^\mu \frac{\delta W}{\delta u_\gamma ^\mu (x)}=0
\eqnum{3.12}
\end{equation}
With introduction of the generating functional of proper vertices defined by 
\begin{equation}
\Gamma [\Phi ,K]=W[J,K]-\int d^4xJ_i(x)\Phi _i(x)  \eqnum{3.13}
\end{equation}
where $\Phi _i$ are the vacuum expectation values of the field operators in
the presence of external sources, one may easily write down the
representations of the identity in Eq.(3.6) and the ghost equations in
Eqs.(3.9)-(3.12) through the functional $\Gamma $ which we do not list here.

\section{Inclusion of quarks}

In the previous sections, the quantum electroweak theory for leptons has
been built up starting from the Lagrangian given in the unitary gauge. For
completeness, in this section, the corresponding theory for quarks will be
briefly formulated. The $SU(2)\times U(1)$ symmetric Lagrangian describing
the interactions of quarks with the gauge bosons and the Higgs particle is,
in the unitary gauge, of the form$^{[23,24]}$ 
\begin{equation}
\begin{tabular}{l}
${\cal L}_q=\overline{Q}_{jL}i\gamma ^\mu D_\mu Q_{jL}+\overline{U}%
_{jR}i\gamma ^\mu D_\mu U_{jR}+\overline{D}_{jR}^\theta i\gamma ^\mu D_\mu
D_{jR}^\theta $ \\ 
$-\frac 1{\sqrt{2}}f_j(U)[\overline{Q}_{jL}\widetilde{\phi }_0U_{jR}+%
\overline{U}_{jR}\widetilde{\phi }_0^{+}Q_{jL}]$ \\ 
$-\frac 1{\sqrt{2}}f_j(D)[\overline{Q}_{jL}\phi _0D_{jR}^\theta +\overline{D}%
_{jR}^\theta \phi _0^{+}Q_{jL}]$%
\end{tabular}
\eqnum{4.1}
\end{equation}
where the repeated index $j$ $(j=1,2,3)$ which is the label of quark
generation implies summation, 
\begin{equation}
Q_{jL}=\left( 
\begin{array}{c}
U_{jL} \\ 
D_{jL}^\theta
\end{array}
\right)  \eqnum{4.2}
\end{equation}
is the $SU(2)$ doublet (for a given $j$) constructed by the left-handed
quarks in which $U_j$ stands for the up-quark $u,c$ or $t$ and $D_j^\theta $
is defined by 
\begin{equation}
D_j^\theta =V_{jk}D_k,  \eqnum{4.3}
\end{equation}
here $V_{jk}$ denote the elements of the unitary $K-M$ mixing matrix $%
V^{[24,25]}$ and $D_k$ symbolizes the down-quark $d,s$ or $b$, $U_{jR}$ and $%
D_{jR}$ designate the SU(2) singlets for the right-handed up-quarks and
down-quarks respectively, $\phi _0$ is the scalar field doublet defined in
Eq.(2.9), $\widetilde{\phi }_0$ is the charge-conjugate of $\phi _0$ which
is defined by$^{[24]}$ 
\begin{equation}
\widetilde{\phi }_0=i\tau _2\phi _0^{*}=\left( 
\begin{array}{c}
H+v \\ 
0
\end{array}
\right)  \eqnum{4.4}
\end{equation}
$f_j(U)$ and $f_j(D)$ are the coupling constants.

In Eq.(4.1), the first three terms are responsible for determining the
kinetic energy terms of quarks and the interactions between the quarks and
the gauge bosons, and the remaining terms which are simpler than those
chosen in the $R_\alpha $-gauge theory are designed to yield the quark
masses and the couplings between the quarks and the Higgs particle. By using
the expressions shown in Eqs.(4.2)-(4.4) and their conjugate ones as well as
the following eigen-equations 
\begin{eqnarray}
YQ_{jL} &=&\frac 13Q_{jL}\text{ , }YU_{jR}=\frac 43U_{jR\text{ }}\text{ , }%
YD_{jR}=-\frac 23D_{jR}\text{ ,}  \nonumber \\
\tau ^aU_{jR} &=&0\text{ , }\tau ^aD_{jR}=0  \eqnum{4.5}
\end{eqnarray}
the Lagrangian in Eq.(4.1) will be represented as 
\begin{equation}
\begin{tabular}{l}
${\cal L}_q=\overline{U}_j[i\gamma ^\mu \partial _\mu -m_j\left( U\right)
]U_j+\overline{D}_j[i\gamma ^\mu \partial _\mu -m_j\left( D\right) ]D_j$ \\ 
$+\frac g{\sqrt{2}}[\widetilde{j}_{L\mu }^{-}W^{+\mu }+\widetilde{j}_{L\mu
}^{+}W^{-\mu }]+e\widetilde{j}_\mu ^{em}A^\mu +\frac e{\sin 2\theta _w}%
\widetilde{j}_\mu ^0Z^\mu $ \\ 
$-\frac 1{\sqrt{2}}f_j(U)\overline{U}_jU_jH-\frac 1{\sqrt{2}}f_j\left(
D\right) \overline{D}_jD_jH$%
\end{tabular}
\eqnum{4.6}
\end{equation}
where 
\begin{equation}
\widetilde{j}_{L\mu }^{-}=\overline{U}_{jL}\gamma _\mu V_{jk}D_{kL}=%
\widetilde{(j}_{L\mu }^{+})^{+}  \eqnum{4.7}
\end{equation}
\begin{equation}
\widetilde{j}_\mu ^{em}=\frac 23\overline{U}_j\gamma _\mu U_j-\frac 13%
\overline{D}_ji\gamma _\mu D_j  \eqnum{4.8}
\end{equation}
\begin{equation}
\widetilde{j}_\mu ^0=\overline{U}_j\gamma _\mu \frac 12(1-\gamma _5)U_j-%
\overline{D}_j\gamma _\mu \frac 12(1-\gamma _5)D_j-2\sin ^2\theta _w%
\widetilde{j}_\mu ^{em}  \eqnum{4.9}
\end{equation}
and 
\begin{equation}
m_j(U)=\frac 1{\sqrt{2}}f_j(U)v,m_j(D)=\frac 1{\sqrt{2}}f_j(D)v  \eqnum{4.10}
\end{equation}
From the procedure of quantization as stated in Sect.2, it is clear to see
that the Lagrangian in Eq.(4.1) or (4.6), as a part of the total Lagrangian
of the lepton-quark system, may simply be added to the effective Lagrangian
denoted in Eq.(2.34).

\section{The theory without Higgs boson}

In the previous sections, we described the theory given in the unitary gauge
which includes the Higgs boson in it. Whether the Higgs particle really
exists in the world or not nowadays becomes a central problem in particle
physics. If the Higgs particle could not be found in experiment, one may ask
whether this particle can be thrown out from the electroweak theory without
breaking the original gauge-symmetry of the theory? Recall that the gauge
transformation does not alter the vector nature of the gauge boson fields
and the spinor character of the fermion fields. They all remain the same
numbers of components before and after gauge transformations. But, the
situation for the scalar field is different. In the functional space spanned
by the four scalar functions, the scalar function $\phi $ defined in
Eq.(1.4) forms a four-dimensional vector and the gauge transformations act
as rotations. A special rotation ( the so-called Higgs transformation) can
convert a four-dimensional vector $\phi $ to the one which has only one
nonvanishing component along the Higgs direction, i.e. the function $H(x)+v$%
. But, any rotation does not change the magnitude of the vector $\phi
,\left| \phi \right| =H+v.$ In the Higgs mechanism, although the vacuum
state can be chosen in different ways, it is usually chosen in the Higgs
direction. This choice is made just from the physical requirement. The
scalar field function $H(x)+v$ was viewed as physical. But, the theory does
not tell us what the vacuum expectation value $v$ should be. It may be very
large or very small. The scalar fields and their vacuum expectation value
were introduced originally for giving some gauge bosons and fermions masses
. For this purpose, we may simply limit ourself to require the magnitude of
the vector $\phi $ to be equal to the vacuum expectation $v$, $\left| \phi
\right| =v.$ In this case, we may set $\phi \rightarrow V$ where $V$ was
represented in Eq.(1.6). The vacuum $V$ generally is a function of
space-time; but in practice, as usual, it is chosen to be a constant along
the Higgs direction. With this choice, the Lagrangian of the system under
consideration is still represented by Eq.(2.1) except that the Lagrangian $%
{\cal L}_\phi $ in Eq.(2.8) is reduced to 
\begin{equation}
{\cal L}_\phi ={\cal L}_{GM}+{\cal L}_{lm}  \eqnum{5.1}
\end{equation}
\begin{equation}
\begin{tabular}{l}
${\cal L}_{GM}=(D^\mu V)^{+}(D_\mu V)-\mu ^2V^{+}V-\lambda (V^{+}V)^2$ \\ 
$=\frac 12F_\mu ^{+}F^\mu $%
\end{tabular}
\eqnum{5.2}
\end{equation}
where 
\begin{equation}
F_\mu =\frac 12(g\tau ^aA_\mu ^a+g^{\prime }B_\mu )V  \eqnum{5.3}
\end{equation}
and 
\begin{equation}
{\cal L}_{lm}=-f_l(\overline{L}Vl_R+\overline{l}_RV^{+}L)  \eqnum{5.4}
\end{equation}
The ${\cal L}_{GM}$ and ${\cal L}_{lm}$ now only play the role of generating
the mass terms of W$^{\pm }$ and Z$^{0\text{ }}$ bosons and charged fermions
respectively. Correspondingly, the gauge transformation in Eq.(2.13) is
reduced to 
\begin{equation}
\delta V=\frac i2(g\tau ^a\theta ^a+g^{\prime }\theta ^0)V  \eqnum{5.5}
\end{equation}
here the gauge transformation of $V$ , as pointed out before, is chosen to
be the same as for the scalar function $\phi $. The other gauge
transformations are still represented in Eqs.(2.14)-(2.16 ). It is easy to
see that the Lagrangian ${\cal L}_{lm}$ is gauge-invariant, $\delta {\cal L}%
_{lm}=0$. By the gauge transformations written in Eqs.(5.6) and (2.14)-(2.16
), one may find 
\begin{equation}
\begin{tabular}{l}
$\delta F_\mu =\frac i2(g\tau ^a\theta ^a+g^{\prime }\theta ^0)F_\mu
+\partial _\mu (g\tau ^a\theta ^a+g^{\prime }\theta ^0)V$ \\ 
$\delta F_\mu ^{+}=-\frac i2F_\mu ^{+}(g\tau ^a\theta ^a+g^{\prime }\theta
^0)+V^{+}\partial _\mu (g\tau ^a\theta ^a+g^{\prime }\theta ^0)$%
\end{tabular}
\eqnum{5.6}
\end{equation}
With these transformations, it can be proved that the action given by the
Lagrangian ${\cal L}_{GM}$ and hence the action given by the total
Lagrangian ${\cal L}$ is gauge-invariant under the Lorentz condition written
in Eqs.(1.8) and (1.9), 
\begin{equation}
\begin{tabular}{l}
$\delta S=\int d^4x\delta {\cal L=}\int d^4x\delta {\cal L}_{GM}{\cal =}%
\frac 12\int d^4x(\delta F_\mu ^{+}F^\mu +F_\mu ^{+}\delta F^\mu )$ \\ 
$=-\frac{v^2}4\int d^4x\{g^2(\partial ^\mu A_\mu ^1\theta ^1+\partial ^\mu
A_\mu ^2\theta ^2)+(g^{\prime }\theta ^0-g\theta ^3)(g^{\prime }\partial
^\mu B_\mu -g\partial ^\mu A_\mu ^3)\}$ \\ 
$=0$%
\end{tabular}
\eqnum{5.7}
\end{equation}
As emphasized in Ref.[1] and in Introduction, for a constrained system such
as the massive gauge field, whether a system is gauge-invariant or not
should be seen from that whether the action of the system, which is given in
the physical subspace defined by the introduced constraint conditions, is
gauge-invariant or not. Thus, according to Eq.(5,7), the Lagrangian given in
Eq.(2.1) with the ${\cal L}_\phi $ given in Eqs.(5.1)-(5.4) still ensures
the theory to have the $SU(2)\times U(1$) gauge-symmetry.

Now, let us to quantize the theory by means of the Lagrange multiplier
method. For this purpose, the Lorentz conditions in Eq. (2.12) are extended
to 
\begin{equation}
\partial ^\mu A_\mu ^i+\alpha \lambda ^i=0  \eqnum{5.8}
\end{equation}
where $i=0,1,2,3$ (Note : the conditions In Eqs.(2.10) and (2.11) are not
necessarily considered here because they become trivial identities in this
case)$.$ When these conditions are incorporated into the Lagrangian by the
Lagrange multiplier method, we have 
\begin{equation}
{\cal L}_\lambda ={\cal L+\lambda }^i\partial ^\mu A_\mu ^i+\frac 12\alpha
(\lambda ^i)^2  \eqnum{5.9}
\end{equation}
where ${\cal L}$ was written in Eqs.(2.1)-(2.7) and (5.1)-(5.4). As
mentioned before, in order to build a gauge-invariant theory, the action
given by the above Lagrangian must be required to be gauge-invariant. By
making use of the gauge transformation in Eq.(2.16) and the gauge
transformation of the Lagrangian ${\cal L}$ which may be read from Eq.(5.7), 
\begin{equation}
\begin{tabular}{l}
$\delta {\cal L}=-\frac{v^2}4\{g^2(\partial ^\mu A_\mu ^1\theta ^1+\partial
^\mu A_\mu ^2\theta ^2)+(g^{\prime }\theta ^0-g\theta ^3)(g^{\prime
}\partial ^\mu B_\mu -g\partial ^\mu A_\mu ^3)\}$ \\ 
$=-m_W^2\partial ^\mu W_\mu ^{+}\theta ^{-}-m_W^2\partial ^\mu W_\mu
^{-}\theta ^{+}-m_Z^2\partial ^\mu Z_\mu \theta ^0$%
\end{tabular}
\eqnum{5.10}
\end{equation}
where we have used the definitions denoted in Eqs.(2.29)-(2.33) and (2.37)
and utilizing the constraint condition in Eq.(5.8), one can derive 
\begin{equation}
\begin{tabular}{l}
$\delta S=\int d^4x\{\delta {\cal L-}\frac 1\alpha \partial ^\mu A_\mu
^i\partial ^\nu \delta A_\nu ^i\}$ \\ 
$=-\frac 1\alpha \{\partial ^\nu W_\nu ^{+}(\partial ^\mu \delta W_\mu
^{-}+\alpha m_W^2\theta ^{-})+\partial ^\nu W_\nu ^{-}(\partial ^\mu \delta
W_\mu ^{+}+\alpha m_W^2\theta ^{+})$ \\ 
$-\partial ^\nu Z_\nu (\partial ^\mu \delta Z_\mu +\alpha m_Z^2\theta
^0)-\partial ^\nu A_\nu \partial ^\mu \delta A_\mu \}$ \\ 
$=0$%
\end{tabular}
\eqnum{5.11}
\end{equation}
According to Eq.(5.8), $\frac 1\alpha \partial ^\nu W_\nu ^{\pm }\neq 0,%
\frac 1\alpha \partial ^\nu Z_\nu \neq 0$ and $\frac 1\alpha \partial ^\nu
A_\nu \neq 0$, therefore, we may obtain from Eq.(5.11) the constraint
equations on the gauge group 
\begin{equation}
\partial ^\mu \delta W_\mu ^{\pm }+\alpha m_W^2\theta ^{\pm }=0  \eqnum{5.12}
\end{equation}
\begin{equation}
\partial ^\mu \delta Z_\mu +\alpha m_Z^2\theta ^0=0  \eqnum{5.13}
\end{equation}
\begin{equation}
\partial ^\mu \delta A_\mu =0  \eqnum{5.14}
\end{equation}
Defining the ghost field functions $C^i$ by $\theta ^i=\xi C^i$ and omitting
the infinitesimal anticommuting number $\xi $ from Eqs.(5.12)-(5.14), we get
the ghost equations as follows 
\begin{equation}
\partial ^\mu \Delta W_\mu ^{\pm }+\alpha m_W^2C^{\pm }=0  \eqnum{5.15}
\end{equation}
\begin{equation}
\partial ^\mu \Delta Z_\mu +\alpha m_Z^2C^0=0  \eqnum{5.16}
\end{equation}
\begin{equation}
\partial ^\mu \Delta A_\mu =0  \eqnum{5.17}
\end{equation}
where $\Delta W^{\pm },\Delta Z$ and $\Delta A_\mu $ were defined in
Eq.(3.2). The constraint equations in Eqs.(5.15)-(5.17) may also be
incorporated into the Lagrangian in Eq.(5.9) by the Lagrange undetermined
multiplier method to give a generalized Lagrangian 
\begin{equation}
{\cal L}_\lambda ^{*}={\cal L+\lambda }^i\partial ^\mu A_\mu ^i+\frac 12%
\alpha (\lambda ^i)^2+{\cal L}_{gh}  \eqnum{5.18}
\end{equation}
where 
\begin{equation}
\begin{tabular}{l}
${\cal L}_{gh}=\overline{C}^{+}(\partial ^\mu \Delta W_\mu ^{-}+\alpha
m_W^2C^{-})+\overline{C}^{-}(\partial ^\mu \Delta W_\mu ^{+}+\alpha
m_W^2C^{+})+\overline{C}_Z(\partial ^\mu \Delta Z_\mu +\alpha m_Z^2C^0)+%
\overline{C}_\gamma \partial ^\mu \Delta A_\mu $ \\ 
$=\overline{C}^{-}(\Box +\alpha M_w^2)C^{+}+\overline{C}^{+}(\Box +\alpha
M_w^2)C^{-}+\overline{C}_z(\Box +\alpha M_z^2)C_z+\overline{C}_\gamma \Box
C\gamma -ig\{(\partial ^\mu \overline{C}^{+}C^{-}$ \\ 
$-\partial ^\mu \overline{C}^{-}C^{+})(\cos \theta _wZ_\mu +\sin \theta
_wA_\mu )+(\partial ^\mu \overline{C}^{-}W_\mu ^{+}-\partial ^\mu \overline{C%
}^{+}W_\mu ^{-})(\cos \theta _wC_z+\sin \theta _wC_\gamma )$ \\ 
$+(\cos \theta _w\partial ^\mu \overline{C}_z+\sin \theta _w\partial ^\mu 
\overline{C}_\gamma )(C^{+}W_\mu ^{-}-C^{-}W_\mu ^{+})\}$%
\end{tabular}
\eqnum{5.19}
\end{equation}
which is simpler than that given in Eq.(2.45).

Completely following the procedure shown in Eq.(2.53), we may use the
Lagrangian in Eq.(5.18) to construct the generating functional of Green's
functions which is formally as the same as that in Eq.(2.53). After
calculating the integral over $\lambda ^i$, we obtain an effective
Lagrangian like this 
\begin{equation}
{\cal L}_{eff}={\cal L}_G+{\cal L}_F++{\cal L}_{gf}+{\cal L}_{gh}+{\cal L}_q
\eqnum{5.20}
\end{equation}
where ${\cal L}_G,{\cal L}_F$ and ${\cal L}_{gf}$ were represented in
Eqs.(2.34)-(2.37), (2.38)-(2.42) and (2.44) respectively, while, ${\cal L}%
_{gh}$ is now given in Eq.(5.19) and ${\cal L}_q$ is still described in
section 4 except that the scalar fields $\phi _0$ and $\widetilde{\phi }_0$
in Eq.(4.1) are now replaced by the vacuum $V$ denoted in Eq.(1.6) and,
therefore, the last two terms in Eq.(4.6) are absent in the present case.
The results stated above can equally be derived by employing the
Faddeev-Popov approach. The Lagrangian given in this section establishes the
quantum electroweak theory without involving the Higgs boson. This theory
may actually be written out from the theory described in sections 2, 3 and 4
by dropping out all the terms related to the Higgs boson.

\section{Remarks}

The quantum electroweak theory described in the previous sections is not
only simpler than the ordinary $R_\alpha $-gauge theory, but also would
safely ensure the theory to be renormalizable due to the absence of the
Goldstone bosons. To this end, we would like to mention the role played by
the Goldstone bosons in the ordinary theory. As illustrated by the example
presented in Appendix A which shows the tree diagrams of
antineutrino-electron scattering and their S-matrix elements, the Goldstone
boson propagator in Fig.(b) just plays the role of cancelling out the
contribution arising from the unphysical part of the gauge boson propagator
in Fig.(a) to the S-matrix element. Therefore, the ordinary R$_\alpha $%
-gauge theory can naturally guarantee the tree unitarity of the S-matrix
element. However, considering that the both diagrams in Figs.(a) and (b), as
subgraphs, will appear, companying each other, in higher order Feynman
diagrams and they can be replaced by the only one diagram shown in Fig.(a)
in which the gauge boson propagator is given in the unitary gauge, the bad
ultraviolet divergence of the term in the latter propagator would cause some
difficulties of renormalization as indicated in Ref.[16]. In contrast, in
the theory presented in this paper, there are not the Feynman diagrams
involving the Goldstone bosons like Fig.(b), therefore, the aforementioned
term of bad ultraviolet behavior does not appear in the massive gauge boson
propagator and any Feynman integrals to spoil the renormalizability of the
theory. However, the present theory formulated in the $\alpha $-gauge will
not content with the unitarity condition of S-matrix elements due to the
presence of the axial current and the mass difference between the charged
particle and neutral one. How to understand and resolve this problem? As
explained in Appendix B, the propagator in Eq.(1.1) is given by the physical
transverse vector potential which is on the mass-shell (this point is
clearly seen in the canonical quantization; but not so clearly in the
path-integral quantization), While, the propagator in Eq.(1.6) is given by
the full vector potential which contains an unphysical longitudinal
component in it and therefore is off-mass-shell. This propagator is suitable
to be used for calculating Green's functions which are off-shell. In the
limit: $\alpha \rightarrow \infty ,$ the $\alpha $-gauge propagator is
converted to the unitary gauge one since the unphysical part of the former
propagator vanishes in the limit. Therefore, the propagator given in the $%
\alpha $ -gauge can be considered as a kind of parametrization (or say,
regularization) of the propagator given in the unitary gauge, somehow
similar to the regularization procedure in the renormalization scheme. In
view of this point of view, we have no reasons to require the $\alpha $%
-gauge theory to directly give the on-shell S-matrix elements. Nevertheless,
due to its renormalizable character, it is suitable to use such a theory at
first in practical calculations of the S-matrix elements and then the $%
\alpha -$limiting procedure mentioned above is necessary to be required in
the final step of the calculations$^{[18]}$. It is noted that the $\alpha -$%
limiting procedure can only be applied to the massive gauge boson
propagators. This means that we have to make distinction between the gauge
parameters appearing in the massive gauge boson propagators and the photon
propagator in the procedure. Certainly, by the limiting procedure, the
unitarity of the theory is always ensured in spite of whether the currents
involved in the theory are conserved or not. At last, we mention that the
ordinary $R_\alpha $ -gauge theory, actually, can also be viewed as another
kind of parametrization of the unitary gauge theory because in the limit: $%
\alpha \rightarrow \infty ,$ the theory in the R$_\alpha -$gauge directly
goes over to the one in the unitary gauge. The question arises: which
parametrization is suitable? The answer should be given by the requirement
that which theory allows us to perform the renormalization safely and give
correct physical results. An essential point to fulfil this requirement is
that the theory must maintain the original gauge-symmetry, just as the same
requirement for the regularization procedure of renormalization. The $\alpha 
$-gauge theory formulated in this paper is exactly of the $SU(2)\times U(1)$
gauge symmetry. As shown in Sect.3, this gauge symmetry is embodied in the
W-T identities satisfied by the generating functionals. From these W-T
identities, one may readily derive a set of W-T identities obeyed by Green's
functions and vertices which establish correct relations between the Green's
functions and the vertices and provide a firm basis for performing the
renormalization of the theory. These subjects will be discussed in the
subsequent papers.

\section{ A{\bf cknowledgment}}

This project was supported in part by National Natural Science Foundation of
China.

\section{\bf Appendix A: The lowest-order S-matrix element of
antineutrino-electron scattering}

The tree diagrams representing the antineutrino-electron scattering are
shown in Figs.(a)- (c) in which the internal lines are respectively the
W-boson propagator, the Goldstone boson propagator and the Z$^0$ boson
propagator. According to the Feynman rules given in the $R_\alpha $-gauge
theory$^{[7,23]}$, the corresponding S-matrix element can be written as 
\begin{equation}
T_{fi}=T_w+T_G+T_Z  \eqnum{A.1}
\end{equation}
where $T_w,T_{G\text{ }}$and $T_Z$ represent respectively the S-matrix
elements of Figs.(a), (b) and (c), 
\begin{equation}
\begin{tabular}{l}
$T_w=\left( \frac{-ig}{2\sqrt{2}}\right) ^2\overline{u}_e(q_1)\gamma ^\mu
(1-\gamma _5)v_\nu (q_2)\overline{v}_\nu (p_2)\gamma ^\nu (1-\gamma
_5)u_e(p_1)$ \\ 
$\times \frac{-i}{k^2-M_w^2+i\varepsilon }[g_{\mu \nu }-(1-\alpha )\frac{%
k_\mu k_\nu }{k^2-\alpha M_w^2}]$%
\end{tabular}
\eqnum{A.2}
\end{equation}
\begin{equation}
T_G=\left( \frac{if_e}2\right) ^2\overline{u}_e(q_1)(1-\gamma _5)v_\nu (q_2)%
\overline{v}_\nu (p_2)(1+\gamma _5)u_e(p_1)\frac i{k^2-\alpha M_w^2} 
\eqnum{A.3}
\end{equation}
where $k=p_1+p_2=q_1+q_2$ and the subscripts ''$e$'' and ''$\nu $'' mark
which particles the spinors belong to, and 
\begin{equation}
\begin{tabular}{l}
$T_Z=\frac{g^2}{8\cos ^2\theta _w}\overline{u}_e(q_1)\gamma ^\mu [\frac 12%
(1-\gamma _5)-2\sin ^2\theta _w]u_e(p_1)\overline{v}_\nu (p_2)\gamma ^\nu
(1-\gamma _5)v_\nu (q_2)$ \\ 
$\times \frac{-i}{k^2-M_Z^2+i\varepsilon }[g_{\mu \nu }-(1-\alpha )\frac{%
k_\mu k_\nu }{k^2-\alpha M_Z^2}]$%
\end{tabular}
\eqnum{A.4}
\end{equation}
where $k=q_1-p_1=p_2-q_2.$

Noticing the decomposition 
\begin{equation}
g_{\mu \nu }-\left( 1-\xi \right) \frac{k_\mu k_\nu }{k^2-\alpha M_j^2}%
=g_{\mu \nu }-\frac{k_\mu k_\nu }{M_j^2}+\frac{\left( k^2-M_j^2\right) k_\mu
k_\nu }{\left( k^2-\alpha M_j^2\right) M_j^2}  \eqnum{A.5}
\end{equation}
where $j=W$ or $Z$, the matrix element in Eq.(A.2) can be rewritten as 
\begin{equation}
T_w=T_w^{(1)}+T_w^{(2)}  \eqnum{A.6}
\end{equation}
where 
\begin{equation}
T_w^{(1)}=\left( \frac{-ig}{2\sqrt{2}}\right) ^2\overline{u}_e(q_1)\gamma
^\mu (1-\gamma _5)v_\nu (q_2)\overline{v}_\nu (p_2)\gamma ^\nu (1-\gamma
_5)u_e(p_1)\frac{-i}{k^2-M_w^2}(g_{\mu \nu }-\frac{k_\mu k_\nu }{M_w^2}) 
\eqnum{A.7}
\end{equation}
and 
\begin{equation}
T_w^{(2)}=\left( \frac{-ig}{2\sqrt{2}}\right) ^2\overline{u}_e(q_1)\gamma
^\mu (1-\gamma _5)v_\nu (q_2)\overline{v}_\nu (p_2)\gamma ^\nu (1-\gamma
_5)u_e(p_1)\frac{-ik_\mu k_\nu }{M_w^2(k^2-\alpha M_w^2)}  \eqnum{A.8}
\end{equation}
Applying the energy-momentum conservation and the Dirac equation, it is easy
to see

\begin{eqnarray}
&&\ \overline{u}_e(q_1)\gamma ^\mu k_\mu (1-\gamma _5)v_\nu (q_2)\overline{v}%
_\nu (p_2)\gamma ^\nu k_\nu (1-\gamma _5)u_e(p_1)  \nonumber \\
\ &=&m_e^2\overline{u}_e(q_1)(1-\gamma _5)v_\nu (q_2)\overline{v}_\nu
(p_2)(1+\gamma _5)u_e(p_1)  \eqnum{A.9}
\end{eqnarray}
Inserting Eq.(A.9) into Eq.(A8) and using the relation 
\begin{equation}
f_e=\frac{gm_e}{\sqrt{2}M_w}  \eqnum{A.10}
\end{equation}
we find 
\begin{equation}
T_w^{(2)}=-T_G  \eqnum{A.11}
\end{equation}
Therefore, we have 
\begin{equation}
T_w+T_G=T_w^{(1)}  \eqnum{A.12}
\end{equation}
This result shows us that the contributions from Figs.(a) and (b) is equal
to the contribution from Fig.(a) provided that the W-boson propagator in
Fig.(a) is replaced by the one given in the unitary gauge. Similarly,
employing Eq.(A.5) and the Dirac equations for electron and neutrino,
Eq.(A.4) becomes 
\begin{equation}
\begin{tabular}{l}
$T_Z=\frac{g^2}{8\cos ^2\theta _w}\overline{u}_e(q_1)\gamma ^\mu [\frac 12%
(1-\gamma _5)-2\sin ^2\theta _w]u_e(p_1)\overline{v}_\nu (p_2)\gamma ^\nu
(1-\gamma _5)v_\nu (q_2)$ \\ 
$\times \frac{-i}{k^2-M_Z^2+i\varepsilon }[g_{\mu \nu }-\frac{k_\mu k_\nu }{%
k^2-M_Z^2}]$%
\end{tabular}
\eqnum{A.13}
\end{equation}
where the Z$^0$ boson propagator is also given in the unitary gauge.

\section{Appendix B: On the massive gauge boson propagators}

To help understanding of the nature of the massive gauge boson propagators
given in the unitary gauge and the $\alpha $ -gauge, we show how these
propagators are derived in the formalism of canonical quantization. For
simplicity, we only take the Lagrangian of a free massive vector field$%
^{[22]}$

\begin{equation}
{\cal L}_0{\frak =-}\frac 14F^{\mu \nu }F_{\mu \nu }+\frac 12m^2V^\mu V_\mu 
\eqnum{B.1}
\end{equation}
where $V_\mu $ is the vector potential for a massive vector field (W$^{\pm }$
or Z$^0$). To give a complete formulation of the field dynamics, the above
Lagrangian must be constrained by the Lorentz condition: $\partial ^\mu
V_\mu =0$ whose solution is $V_{L\mu }=0.$ Substituting this solution in
Eq.(B.1), the Lagrangian will be merely expressed by the transverse vector
potential $V_{T\mu }.$ $%
\mathop{\rm Si}
$nce the $V_{T\mu }$ completely describes the three independent polarization
states of the massive vector field, in operator formalism, it can be
represented by the following Fourier integral 
\begin{equation}
V_{T\mu }(x)=\int \frac{d^3k}{(2\pi )^{3/2}}\sum_{\lambda =1}^3\frac{%
\epsilon _\mu ^\lambda ({\bf k})}{\sqrt{2\omega ({\bf k)}}}[a_\lambda ({\bf k%
})e^{-ikx}+a_\lambda ^{+}({\bf k})e^{ikx}]  \eqnum{B.2}
\end{equation}
where $\omega ({\bf k)}$ is the energy of free particle and $\epsilon _\mu
^\lambda ({\bf k})$ is the unit vector of polarization satisfying the
transversity condition: $k^\mu \epsilon _\mu ^\lambda ({\bf k})=0,$ which
corresponds to the transversity condition: $\partial ^\mu V_{T\mu }(x)=0.$
By using the familiar canonical commutation relations between the
annihilation operator $a_\lambda ({\bf k})$ and the creation one $a_\lambda
^{+}({\bf k})$, as derived in the literature$^{[22]}$, one gets the
propagator for the transverse vector potential as follows 
\begin{equation}
iD_{\mu \nu }(x-y)=\left\langle 0\left| T\{V_{T\mu }(x)V_{T\nu }(y)\}\right|
0\right\rangle =\int \frac{d^4k}{(2\pi )^4}iD_{\mu \nu }(k)e^{-ik(x-y)} 
\eqnum{B.3}
\end{equation}
where $iD_{\mu \nu }(k)$ is just the one shown in Eq.(1.1) here a
non-covariant part of the propagator has been omitted because it will be
cancelled in S-matrix elements by the non-covariant term in the interaction
Hamiltonian .

On the other hand, when the Lorentz condition is generalized to the from: $%
\partial ^\mu V_\mu +\alpha \lambda =0,$ where $\lambda $ acts as a Lagrange
multiplier, and incorporated into the Lagrangian by the Lagrangian
multiplier method, one may obtain the St\"uckelberg's Lagrangian$^{[22]}$ 
\begin{equation}
{\cal L}_0{\frak =-}\frac 14F^{\mu \nu }F_{\mu \nu }+\frac 12m^2V^\mu V_\mu -%
\frac 1{2\alpha }(\partial ^\mu V_\mu )^2  \eqnum{B.4}
\end{equation}
According to the spirit of the Lagrange multiplier method, every component
of the vector potential in the above Lagrangian can be treated as
independent. From the Lagrangian, one may derive the equation of motion 
\begin{equation}
\left( \Box +m^2\right) V_\mu -(1-\frac 1\alpha )\partial _\mu \partial _\nu
V^\nu =0  \eqnum{B.5}
\end{equation}
Taking divergence of the above equation leads to a scalar field equation 
\begin{equation}
(\Box +\mu ^2)\varphi =0  \eqnum{B.6}
\end{equation}
where $\varphi =\partial ^\mu V_\mu $ and $\mu ^2=\alpha m^2$. Now the full
vector potential can be expressed as 
\begin{equation}
V_\mu =V_{T\mu }+V_{L\mu }  \eqnum{B.7}
\end{equation}
where $V_{T\mu }$ is the transverse part of the potential which was
represented in Eq.(B.2) and $V_{L\mu }$ is the longitudinal part of the
potential which is defined by $V_{L\mu }=\frac 1{\mu ^2}\partial _\mu
\varphi $ and can be expanded as 
\begin{equation}
V_{L\mu }(x)=\int \frac{d^3k}{(2\pi )^{3/2}}\frac{\epsilon _\mu ^0(k{\bf )}}{%
\sqrt{2\varpi }}[a_0({\bf k})e^{-ikx}+a_0^{+}({\bf k})e^{ikx}]  \eqnum{B.8}
\end{equation}
where $\epsilon _\mu ^0(k)=k_\mu /m$ and $\varpi $ is the energy of the
scalar particle of mass $\mu .$ With the expressions presented in Eqs.(B.7),
(B.2) and (B.8), in the canonical formalism, it is easy to derive the
propagator for the full vector potential$^{[22]}$ 
\begin{equation}
iD_{\mu \nu }(x-y)=\left\langle 0\left| T\{V_\mu (x)V_\nu (y)\}\right|
0\right\rangle =\int \frac{d^4k}{(2\pi )^4}iD_{\mu \nu }(k)e^{-ik(x-y)} 
\eqnum{B.9}
\end{equation}
where $iD_{\mu \nu }(k)$ is exactly of the form as written in Eq.(1.7). This
propagator is usually derived in the path-integral formalism and often used
in perturbative calculations. The propagator in Eq.(1.7) contains two parts:
The first part which is usually given in the Landau gauge ($\alpha =0$) is
transverse with respect to the off-shell momentum $k_\mu $, while the second
part is longitudinal for the off-shell momentum. According to the
decomposition in Eq.(A.5), the above propagator can be divided into such two
parts: one is that given in the unitary gauge as shown in Eq.(1.1); another
is 
\begin{equation}
iD_{\mu \nu }(k)=-i\frac{k_\mu k_\nu }{\left( k^2-\alpha M^2\right) M^2} 
\eqnum{B.10}
\end{equation}
The two parts are respectively transverse and longitudinal with respect to
the on-shell momentum $k_\mu .$ As one can see, the off-shell transverse
propagator (in the Landau gauge) and the on-shell-transverse propagator (in
the unitary gauge) are given by different limits: $\alpha \rightarrow 0$ and 
$\alpha \rightarrow \infty ,$ respectively.

\section{\bf References}

[1] J. C. Su, Nuovo Cimento 117B (2002) 203.

[2] L. D. Faddeev and V. N. Popov, Phys. Lett. B25, 29 (1967); L. D. Faddeev
and A. A. Slavnov,

Gauge Fields: Introduction to Quantum Theory, The Benjamin Commings
Publishing Company Inc. (1980).

[3] J. C. Su, hep.th/9805192; 9805193; 9805194.

[4] S. L. Glashow, Nucl. Phys. 22, 579 (1961).

[5] S. Weinberg, Phys. Rev. Lett. 19, 1264 (1967).

[6] A. Salam, Elementary Particle Theory, ed. by N. Svartholm, Almqvist and
Wiksell, Stockholm, 367 (1968).

[7] K. Fujikawa, B. W. Lee and A. I. Sanda, Phys.Rev. D6, 2923 (1972).

[8] G. 't Hooft and M. J. G. Veltman, Nucl. Phys. B35, 167 (1971).

[9] G. 't Hooft and M J. G. Veltman, Nucl. Phys. B50, 318 (1972).

[10] B. W. Lee and J. Zinn-Justin, Phys. Rev. D5, 3121, 3137, 3155 (1972);
D7, 1049 (1973).

[11] E. S. Abers and B. W. Lee, Phys. Rep. 9C. 1 (1973).

[12] B. W. Lee, Methods in Field Theory Les Houches, 79 (1974), ed. by R.
Balian and J. Zinn-Justin

(North-Holland).

[13] C. Becchi, A. Rouet and B. Stora, Ann. Phys. 98, 287 (1976).

[14] L. Banlieu and R. Coquereanx, Ann. Phys 140, 163 (1982).

[15] K. Aoki, Z. Hioki, R. Kawabe, M. Konuma and T. Muta, Prog. Theor. Phys.
Suppl. 73, (1982).

[16] Hung Cheng and S. P. Li, hep. th /9901129; hep. th /9902212.

980).

[17] G. t'Hooft, Under the spell of the gauge principle, Advanced Series in
the Mathematical Physics 19 (1994),

Editors: Araki et al. World Scientific, Singapore.

[18] T. D. Lee and C. N. Yang, Phys. Rev. 128, 885 (1962).

[19] I. Bars and M. Yoshimura, Phys. Rev. D6, 374 (1972).

[20] J. C. Ward, Phys. Rev. 78, 182 (1950); Y. Takahashi, Nuovo Cim. 6, 370
(1957).

[21] C. Becchi, A. Rouet and B. Stora. Comm. Math. Phys. 42, 127 (1975).

[22] C. Itzykson and J-B. Zuber, Quantum Field Theory, McGraw-Hill Inc.
(1980).

[23] M. E. Peskin and D. V. Schroeder, An Introduction to Quantum Field
Theory, Addison-Wesley Publishing

Company, (1995).

[24] S. L. Glashow, J. Iliopoulos and L .Maiani, Phys. Rev. D2, 1285 (1970).

[25] M. Kobayashi and T. Maskawa, Prog. Theor. Phys. 49, 652 (1973).

\section{Figure captions}

Fig.(a): The tree diagram representing the antineutrino-electron scattering
which takes place via the interaction mediated by one W-boson exchange.

Fig.(b): The tree diagram representing the antineutrino-electron scattering
which takes place via the interaction mediated by one Goldstone boson
exchange.

Fig.(c): The tree diagram representing the antineutrino-electron scattering
which takes place via the interaction mediated by one Z$^0$-boson exchange.

\end{document}